\documentclass[a4paper,11pt]{article}
\usepackage{pos_1,bm}

\title{Neutrino propagation in moving and polarized matter}

\author*[a]{Alexander Grigoriev}
\author[b]{Alexander Studenikin}
\author[a]{Alexei Ternov}

\affiliation[a]{Department of Theoretical Physics, Moscow Institute of Physics and Technology,\\
  141701 Moscow Region, Dolgoprudny, Russia}

\affiliation[b]{Department of Theoretical Physics and MSU Branch in Sarov,  Lomonosov Moscow State University, \\
119992 Moscow,  Russia}

\emailAdd{ax.grigoriev@mail.ru}
\emailAdd{studenik@srd.sinp.msu.ru, a-studenik@yandex.ru}
\emailAdd{ternov.ai@mipt.ru}

\abstract{The rapid development of neutrino astronomy, which is expressed, among other things, in the emergence of new neutrino mega-projects capable of efficient registering astrophysical neutrino fluxes requires a detailed knowledge of neutrino evolution inside neutrino sources (type II supernovae, gamma-ray bursts). This evolution can be influenced by many factors, each should be accounted for by a relevant theory. In this work, we develop the theory of neutrino propagation in moving and/or polarized matter by introducing for the first time an exact spin integral of motion. This enables us to obtain the neutrino dispersion under these conditions and resolve it for most important cases. Our approach opens up the possibility to consistently classify neutrino states in moving and/or polarized medium and, as a consequence, to give a systematic description of the related physical phenomena (e.g., neutrino oscillations, neutrino electromagnetic radiation).}

\FullConference{%
  41st International Conference on High Energy physics - ICHEP2022\\
  6-13 July, 2022\\
  Bologna, Italy
}

\begin{document}
\maketitle

An analytical description of neutrino behavior in external matter and fields and specifically that of neutrino spin dynamics not only provides the understanding of the basics of neutrino spin oscillations in astrophysical media but also enables to discover and give account for another new phenomena. Examples include the effects of neutrino Spin Light \cite{StudTern_PLB2005}, and neutrino beam self-polarization in matter \cite{StudTern_PLB2005,LobStud_PLB2004}. All these effects arise due to helicity dependence of neutrino energy in non-moving matter. In this study we go further in developing the neutrino theory in a medium and consider matter motion. In this case the helicity is no longer the exact quantum number and for consistent treatment one has to find the spin integral of motion. Once found, it enables easy derivation of the neutrino dispersion relation and clear physical interpretation of the spin quantum number entered in it.

Our starting point is the modified Dirac equation that describes a neutrino coherently interacting with particles of the external matter, taking into account possible effects of matter motion and polarization \cite{StudTern_PLB2005}
\begin{equation}\label{Dirac_eq}
  \left\{ i\gamma_{\mu}\partial^{\mu}- \frac{1}{2}\gamma_{\mu}(1+\gamma^5)f^{\mu} - m \right\}\Psi(x)=0,
\end{equation}
where $\gamma^{5}=-i\gamma^{0}\gamma^{1}\gamma^{2}\gamma^{3}$ and the 4-vector $f^{\mu}$ is defined once the matter model is chosen, but generally it depends on matter characteristics -- density, velocity and polarization. We will consider homogenous matter, moving in the given reference frame with an arbitrary constant speed $\mathbf{v}$ and having the number density $n=\gamma n_0$, where $n_0$ is the number density in the matter rest frame ($\gamma=1/\sqrt{1-\mathrm{v}^2}$ being the Lorentz gamma factor). We also take the matter to be  composed of electrons (generalization to a more complex composition is just straightforward), and then within the framework of the minimally extended Standard Model we have \cite{StudTern_PLB2005}:
\begin{equation}\label{f_mu}
  \frac{1}{2}\, f^{\mu}=\gamma\tilde{n}_0 (1,\,\mathbf{v})=\tilde{n}_0 v^{\mu}, \quad \tilde{n}_0=\frac{1}{2\sqrt{2}} G_F (1+4\sin^2\theta_W)n_0.
\end{equation}
In order to define the conserved spin quantity one has to find the spin operator which commutes with the Hamiltonian of motion. From Eq.~(\ref{Dirac_eq}) we find:
\begin{equation}\label{H}
  \mathrm{H}=({\bm \alpha}{\mathbf{p}})-{\tilde{n}}({\bm \alpha}{\mathbf{v}})+\tilde{n}({\bm \Sigma}{\mathbf{v}})+\gamma^5\tilde{n}+\tilde{n}+\gamma^0m,
\end{equation}
where we have introduced the neutrino momentum $\mathbf p$ and used the notations $\alpha_i=\gamma^0\gamma^i$, $\Sigma_i=\gamma^0\gamma^5\gamma^i$,  $\tilde{n}=\gamma\tilde{n}_0$. It should be noted that the same structure of the Hamiltonian also covers matter polarization, so that our results below can be easily adopted for that case.

The spin operator is obtained by the following procedure. Let us take the 4-vector spin polarization operator $\mathrm{T}^{\mu}$ (see, for instance, \cite{Sokolov-Ternov-Rel-El}):
\begin{equation}\label{T_mu}
  \mathrm{T}^{\mu}=\gamma^5\gamma^{\mu}-\gamma^5p^{\mu}/m,
\end{equation}
and therein do the momentum ``extension'' $p^{\mu} \rightarrow \tilde{p}^{\mu}\equiv p^{\mu}-f^{\mu}=p^{\mu}-\tilde{n}_0 v^{\mu}$. Denoting the result as $\widetilde{\mathrm{T}}^{\mu}$ we compose its scalar product with the 4-vector $v^{\mu}$: $\widetilde{\mathrm{T}}^{\mu}v_{\mu}\equiv (\widetilde{\mathrm{T}}v)$. Multiplying finally this quantity by $m$ one can verify that the obtained result,
\begin{equation}\label{S}
  \mathrm{S}=m(\widetilde{\mathrm{T}}v),
\end{equation}
commutes with the Hamiltonian (\ref{H}) on solutions of the Dirac equation (i.e. assuming that the energy operator is realized as follows: $p^0 \Psi =E \Psi$, where $E$ is the neutrino energy). Introducing the notation $\widetilde{\mathrm{H}}=\tilde{p}^0=\mathrm{H}-\tilde{n}$, we also present the found spin operator $\mathrm{S}$ in the expanded form:
\begin{equation}\label{S_exp}
  \mathrm{S}=\gamma \left[\gamma^5\gamma^0 m-\gamma^5(\widetilde{\mathrm{H}}-\left(\tilde{{\mathbf{p}}}{\mathbf{v}})\right) -m\gamma^0({\bm \Sigma}{\mathbf{v}}) \right].
\end{equation}
Note that the spin operator (\ref{S}) is a Lorentz scalar. In our study, it is also suitable to introduce another scalar operator $P$:
\begin{equation}\label{P}
  P = (\tilde{p}^{\mu}v_{\mu}) = (\tilde{p}v)=({p}v)-\tilde{n}_0.
\end{equation}
Then, squaring the spin operator $\mathrm{S}$ it is easy to obtain its eigenvalues and write them through $P$:
\begin{equation}\label{Lambda}
   s\sqrt{P^2-m^2} \equiv s\Lambda, \ \ s=\pm 1.
\end{equation}

By this means, the operator $\mathrm{S}$ characterizes the longitudinal particle polarization with respect to the matter velocity 4-vector $v^{\mu}$.  The stationary spin states of the particle are represented by the wave-function $\Psi_{E,s}$, which is an eigenfunction of operators $\mathrm{H}$ and $\mathrm{S}$:
\begin{equation}\label{Eigen_S}
  \mathrm{H}\Psi_{E,s}=E\Psi_{E,s}, \ \ \mathrm{S}\Psi_s=s\Lambda\Psi_s.
\end{equation}
From these equations, it is the standard procedure to derive the neutrino dispersion in which an explicit dependance on the neutrino spin quantum number $s$ is embedded:
\begin{equation}\label{Dispersion}
  p^2-m^2=2\tilde{n}_0(P-s\Lambda),
\end{equation}
where $p^2=E^2-{\mathbf{p}}^2$. This is another form of the neutrino dispersion relation in moving matter, previously found in \cite{PivovStud_2006} (without spin dependance) and \cite{KaloshVoron_2019} (with a polarization variable having another physical meaning).

The obtained dispersion relation (\ref{Dispersion}) enables to easily find neutrino energies for some basic cases. First of all, on putting ${\mathbf v}=0$ and introducing the ``energy sign'' $\varepsilon=\pm 1$ we reproduce the well-known result for neutrino energy in non-moving matter \cite{StudTern_PLB2005}:
\begin{equation}
  (\sqrt{\widetilde{E}^2-m^2}+s\tilde{n}_0)^2=p^2.
\end{equation}
By making the matter speed directed along or against the neutrino momentum ($\mathbf{v} {\parallel} \mathbf{p}$) we have:
\begin{equation}\label{E_v||p}
  E=\varepsilon\sqrt{(\mathrm{p}\mp\mathrm{v}\tilde{n}-s\tilde{n})^2+m^2}\pm s\mathrm{v}\tilde{n}+\tilde{n},
\end{equation}
where the upper signs correspond to the first case and the lower ones to the second.

In the light of the recent interest in the features of neutrino motion in transversal matter currents \cite{Stud_PAN2004}, it also makes sense to write out expression for neutrino energy when $(\mathbf{p v})=0$. In this case, however, Eq.~(\ref{Dispersion}) is the fourth-order equation with respect to energy and we give approximate values for ultrarelativistic neutrino, $\mathrm{p} \gg \tilde{n}$, $\mathrm{p} \gg m$, corresponding to $\varepsilon=1$:
\begin{align}
      E_{s=+1} &=\sqrt{(\mathrm{p}-\tilde{n})^2+m^2}+\tilde{n},\label{1_E_p_perp_v}
      \\
      E_{s=-1} &=\sqrt{\mathrm{p}^2+4\tilde{n}^2\mathrm{v}^2+m^2}+2\tilde{n}.\label{2_E_p_perp_v}
\end{align}
It is interesting to consider in general terms the impact of Eq.~(\ref{2_E_p_perp_v}) on neutrino oscillations $\nu_e \leftrightarrow \nu_\mu$. Associating  $E_{s=-1}$ with left-handed neutrino energy $E_{\nu_L}$ and expanding the root we get
\begin{equation}\label{E_L}
  E_{\nu_L} \simeq \mathrm{p}+2\tilde{n}+\frac{2\tilde{n}^2\mathrm{v}^2}{\mathrm{p}}.
\end{equation}
The second term leads to the standard effective neutrino oscillation potential in matter and the third term provides the shift of this value and consequently the shift of the conventional resonance condition:
\begin{equation}\label{res_cond_change}
  \frac{\Delta}{2\mathrm{p}}\cos 2\theta=\sqrt{2}G_F n + 8\frac{G_F^2 n^2\mathrm{v}^2}{\mathrm{p}}\sin^2\theta_W,
\end{equation}
where the addition is represented by the second term on the right-hand side. The effect of matter motion becomes substantial (when this addiction is of the order of the first term), for example, for $\mathrm{v}=0.9$, $\mathrm{p}\sim 10$~keV and extra-dense matter with $n\sim10^{41}$cm$^{-3}$.

To conclude, we have established the exact neutrino spin integral of motion for the general case of neutrino motion and polarization, determined the corresponding stationary states and obtained the neutrino dispersion relation with the spin quantum number. Besides the purely theoretical significance, these results can be important for the theory of astrophysical neutrino oscillations, for interpreting the data on astrophysical neutrino flux measurement of large-volume detectors as well as for studying various quantum processes with neutrinos in initial and final states in dense astrophysical media.

The work of AS is supported by the Russian Science Foundation under grant No.22-22-00384.


\begin{thebibliography}{99}
\bibitem{StudTern_PLB2005} A. Studenikin, A. Ternov, \emph{Neutrino quantum states and spin light in matter}, \href{https://doi.org/10.1016/j.physletb.2005.01.002}
{\emph{Phys.Lett.B} \textbf{608} (2005) 107} [{\tt hep-ph/0412408}].

\bibitem{LobStud_PLB2004} A. Lobanov, A. Studenikin, \emph{Spin light of neutrino in matter and electromagnetic fields}, \href{https://doi.org/10.1016/S0370-2693(03)00570-7}
{\emph{Phys.Lett.B} \textbf{601} (2004) 171} [{\tt hep-ph/0212393}].

\bibitem{Sokolov-Ternov-Rel-El} A.~A.Sokolov, I.~M.Ternov \emph{Radiation from Relativistic Electrons}, American Institute of Physics, New York 1986.

\bibitem{PivovStud_2006}  I. Pivovarov, A. Studenikin, \emph{Neutrino quantum states in matter}, \href{https://doi.org/10.22323/1.021.0191} {\emph{PoS HEP} \textbf{2005} (2006) 191} [{\tt hep-ph/0512031}].

\bibitem{KaloshVoron_2019} A.\,E. Kaloshin, D.\,M. Voronin, \emph{Neutrino propagation in media and axis of complete polarization}, \href{https://doi.org/10.1140/epjc/s10052-019-6659-x} {\emph{EPJC} \textbf{79} (2019) 153} [{\tt  arXiv:1808.05514}].

\bibitem{Stud_PAN2004} A. Studenikin, \emph{Neutrinos in electromagnetic fields and moving media}, \href{https://doi.org/10.1134/1.1755390} {\emph{Phys.Atom.Nucl.} \textbf{67} (2004) 993}.




\end{thebibliography}
\end{document}